# TripNet: A Method for Constructing Phylogenetic Networks from Triplets

*Ruzbeh Tusserkani*[*], *Changiz Eslahchi*[†], *Hadi Pourmohammadi*[†], *and Azin Azadi*[‡] [⋆]
[*] School of Mathematics, Institute for Research in Fundamental Sciences (IPM). P.O. Box 193955746, Tehran, Iran ; [†] Department of Mathematics, Shahid Beheshti University, G.C. Tehran, Iran; [‡] Department of Mathematical Sciences, Sharif University of Technology, Tehran, Iran; and [⋆] School of Computer Science, Institute for reaserch in Fundamental Sciences (IPM), Tehran, Iran.

We present TripNet, a method for constructing phylogenetic networks from triplets. We will present the motivations behind our approach and its theoretical and empirical justification. To demonstrate the accuracy and efficiency of TripNet, we performed two simulations and also applied the method to five published data sets: Kreitman's data, a set of triplets from real yeast data obtained from the Fungal Biodiversity Center in Utrecht, a collection of 110 highly recombinant Salmonella multi-locus sequence typing sequences, and nrDNA ITS and cpDNA JSA sequence data of New Zealand alpine buttercups of Ranunculus sect. Pseudadonis. Finally, we compare our results with those already obtained by other authors using alternative methods. TripNet, data sets, and supplementary files are freely available for download at (www.bioinf.cs.ipm.ir/softwares/tripnet).

## Introduction

In this paper we present a new algorithm, called TripNet, for constructing phylogenetic networks from a set of triplets. Indeed, TripNet is an algorithm which given an arbitrary (not necessarily dense) set of triplets as input, outputs an adequate phylogenetic network.

Although, due to the importance and the inherent difficulty of the problem of reconstructing phylogenetic networks, making any reasonable attempt to obtain meaningful results could be of potential interest for the phylogenetics community, it is useful to present the rationale behind our method and its theoretical and empirical justifications.

Firstly, in the following four subsections we will explain in detail the rationale of using phylogenetic networks instead of traditional trees to represent evolutionary relationships, our reasons for using triplets as input, the meaning of an " adequate " phylogenetic network, and the importance of working on non-dense sets of triplets.

### Why Networks?

Traditionally, the fundamental task of phylogenetics has been to reconstruct the tree of life using character-based data (Felsenstein, 2004). In the modern molecular phylogenetics, genetic information i.e. inherited molecular sequences stored in the base pair DNA or RNA, are used as characters to construct phylogenetic trees. However, in recent years some new observations have changed our view regarding the output of reconstructing algorithms. As a result, more complex objects, called phylogenetic networks, have emerged as the possible output (Huson et al., 2010).

The practical interest in these objects is twofold: First, there are some evolutionary events like recombination, hybridization, gene conversion, and horizontal gene transfer which all lead to histories that are not adequately modeled by a tree (Linder et al., 2004) , (Sang and Zhong, 2000), (Bordewich and Semple, 2007), (Hallett et al., 2004), (Hudson, 1983), and (Lyngs et al., 2005). Phylogenetic networks are a generalization of phylogenetic trees that permit the representation of non-tree-like underlying histories. Phylogenetic networks which aim at explicitly modeling these types of reticulate evolution are called explicit phylogenetic networks (Huson et al., 2010).

A second reason for interest in phylogenetic networks is the fact that even when the underlying history is treelike, phenomena such as parallel evolution, model heterogeneity, and sampling error can make it difficult to represent the history by a single tree. In such situations networks can provide a useful tool for representing ambiguity or for simultaneously visualizing a collection of feasible trees. Phylogenetic networks which aim at displaying (incompatible) phylogenetic signals are called implicit phylogenetic networks (Huson et al., 2010).

Phylogenetic networks also may arise in novel applications of phylogenetic techniques, for example in the copying history of medieval manuscripts (Bennett et al., 2003).

Up to this point, the efforts to reconstruct phylogenetic networks have mostly focused on building implicit networks and capturing the ambiguities and uncertainties in the input data. TripNet, to the knowledge of the authors of this paper, is the first unsupervised algorithm attempt to infer an explicit phylogenetic network without any biological presumption about the way the reticulate evolution occurs. A key innovation in our method is the use of triplets as the input.

### Why triplets?

In this subsection we explain our reasons for using triplets as input.

a. Filling a Gap : As mentioned before, classical phylogenetics employs character-based algorithms. Since 1980s, and more specifically after the seminal work of Saitou et al., using distance matrices as the input has become very popular (Saitou and Nei, 1987). Moreover, in the past few years quartet-based algorithms have attracted lots of attention from the phylogenetics community (Grunewald et al., 2007), and (Strimmer and Von Haeseler, 1996).

The phylogenetic network methods, similar to phylogenetic tree methods divide into three classes depending on the type of input data. The first class includes methods that construct networks directly from character data. The familiar methods in this class are statistical parsimony (Templeton et al., 1992), and median networks (Bandelt et al., 1995). The second major class of

Key words: Phylogenetic Network, Triplet, Neighbor-Net, QNet, Yeast, Salmonella, Alpine buttercups.
E-mail: ch-eslahchi@sbu.ac.ir

phylogenetic network methods includes those that construct networks from a distance matrix. Neighbor-Net (Bryant and Moulton, 2004) belongs to this class. Finally, the third class includes those that construct network from quartets. QNet (Grunewald et al., 2007) belongs to this class.

Quartet-based algorithms take four-species trees as the input and produce a tree or a network as the result. If we consider character-based methods as an approach using one-species trees and distance-based methods as one dealing with two-species trees, one can recognize that methods incorporating three species trees as a potentially fruitful has been missed in the current studies. TripNet tries to fill this gap.

It is important to note that we are not the first to recognize this gap, as it was previously mentioned by Felsenstein (2004): "It would be possible to estimate three-species trees and then find the full tree as the best possible fit to them. . . . [I]t would be interesting to know whether methods that construct an overall tree from trees of all triples would give noticeably better results than distance methods."

Finally, it is worth mentioning that we are not just filling a gap in theoretical phylogenetics. Using triplets as the input gives us some competitive advantages: overcoming the data disparity problem and the ability to reach the highest accuracy level.

b. Data Disparity problem : Both character-based and distance-based methods take as input a list of sequences that are known for all taxa under study. Therefore, both methods are affected by the well known data disparity problem (Chor, 1998). Often, in a wide range of phylogenetic problems it is impossible to find this common data. Since quartet-based methods can overcome this difficulty, they have established themselves as an important technique in phylogenetics. Since triplet-based methods do not need that common data either, they have the same advantage regarding the data disparity problem as any quartet-based algorithm.

c. Accuracy : Reaching the highest level of accuracy is the most important motivation behind our approach. One of the main challenges in molecular phylogenetics is the managing and mining uncertain data. Errors in sequencing and sequence alignment are found to have a significant negative effect on subsequent inference of phylogenies. There are many sources of error such as imperfections of the current algorithms, computation errors, and also the measurement errors. Thus, one of the most important tasks in molecular phylogenetics would be to clean the data in order to reach accurate data. Otherwise, it would be impossible to have any reliable algorithm.

This brings us to one of the main advantages of TripNet; unweighted triplets are among the most certain inputs. For every triple of sequences $x$, $y$, and $z$ there are three possible triplets: $yz|x$, $zx|y$, or $xy|z$. A triplet such as $yz|x$ contains information that implies that the taxa $y$ and $z$ are closer to each other than to $x$. This information is qualitative and can be extracted with a much higher precision than any quantitative information we might have, like the distance between two taxa or their corresponding sequences. This is due to the fact that we do not need to use all cites in the sequences one by one to build the triplets. Instead, we use the overall information of the sequences for making the triplets so that the resulting data is more accurate and less subject to error. All the outputs produced by TripNet have been made using such brief (but at the same time, very certain) information.

But the triplet-based methods can be valuable if there is some efficient way to construct accurate triplets.

d. The Convenience of Making Triplets : Triplets are easy to construct using the input sequences: all the available softwares, such as PhyML, that construct unrooted binary trees based on the input sequences can be used to produce triplets. It is enough to add an outgroup to all the sets consisting of three sequences in the input and construct a quartet using the current methods. At the end, we can extract the triplet corresponding to each of these groups by removing the outlier sequence. Finally, note that this simple and intuitive method works with a certainty threshold where we have the option to adjust this threshold. Higher thresholds lead to produce only the triplets that we are almost sure about their correctness and leave the rest undecided. Hence, they result in a less dense but more accurate set of triplets as the input. Lower thresholds produce the opposite effect. One advantage of using such methods is the ability to adjust the threshold with respect to our specific needs on the density and the accuracy of the input data.

But even more interesting, in some applications, the data obtained experimentally may already have the form of rooted triplets; for example, Sibley-Ahlquist-style DNA-DNA hybridization experiments can yield rooted triplets directly.

Now suppose that we have constructed a very accurate triplet set from molecular data. If there is no consensus tree for this triplet set then we can conclude with a high degree of confidence that the raw data shows a reticulate evolution. But can we show efficiently that there is no consensus tree for a given triplet set? Rather surprisingly, this question has been answered affirmatively many years ago in a context far different from phylogenetics.

e. The Convenience for Constructing Networks : The most important result about constructing trees from triplets has appeared in the context of database theory. In 1981, Aho et al., studied the problem of constructing a tree from a set of triplets (Aho et al., 1981). They showed that, given a set of triplets, it is possible to construct in polynomial time a rooted tree consistent with all the input triplets, or decide that no such tree exists (Aho et al., 1981). In this paper, we call their algorithm "TripTree". Their method does not construct anything when there is no tree consistent with all the input triplets. But in such cases a simple network may be the answer. TripNet is exactly a method that if the input triplets are not consistent (and consequently, building an exact tree is impossible) then TripNet builds an "adequate" network as close to a tree as possible. TripNet does the same thing in the beginning; if the set of triplets in the input are consistent with a tree, TripNet builds the corresponding exact tree. Otherwise, there must be reticulation nodes with actual biological interpretations.

Which Networks are adequate?

In the future, more reticulate evolution may be found. But currently, reticulate evolution is considered as a “rare nuisance”. So explicit phylogenetic networks do not need to be complex, but do need to reflect biological reality. First, based on the mechanism of evolution of living organisms, we put some restriction on reticulation nodes in every phylogenetic network. For example in a phylogenetic network every reticulation node has indegree 2 and outdegree 1. This means that reticulation nodes are as simple as possible. But among a set of phylogenetic networks, which one is simpler or more tree-like?

To quantify the complexity of a phylogenetic network $N$ two parameter has been considered. The total number of reticulations, $R(N)$, and the maximum number of reticulations in biconnected components, called the level of the network, $L(N)$. Among a set of phylogenetic networks a network with minimum $R(N)$ and/or $L(N)$ is considered to be the most adequate. So there are two different criteria for simplicity of a network depending on which parameter is considered to be the most significant. Recently, the second parameter has received more consideration in the context of phylogenetic networks.

Given an arbitrary number of trees on the same set of taxa, Huynh et al. (2005) describe a polynomial-time algorithm that constructs a level-1 phylogenetic network that displays all trees and has a minimum number of reticulations, if such a network exists. Given a dense triplet set, Jansson and Sung (2006) give a polynomial-time algorithm that constructs a level-1 network consistent with all triplets, if such a network exists.The algorithm by Iersel and Kelk (2009) can be used to find such a network that also minimizes the number of reticulations. These results have later been extended to level-2 (Iersel and Kelk, 2009) and more recently to level-$k$, for all $k$ (To and Habib, 2009).

In Jansson et al. (2007) the authors considered the problem of deciding whether, given a set of triplets as input, it is possible to construct a galled network. They showed that, in general, this problem is $NP$-hard. However, they show that the problem can be solved in polynomial time when the input is dense, that is for each set of three taxa, there is at least one triplet in the input. After their results, all research in this new area has up to this point focused on constructing networks from dense triplet sets.

The motivation behind the first criteria is obvious. Let us explain the motivation for the second criteria. If we consider the reticulate evolutions as events with low probability of occurrence and if we assume that such events occur independently and distributed uniformly over “the tree/network of life”, then we will expect that only a few reticulate evolutions occurs in a biconnected component. So the level of a phylogenetic network must be low. However, from a biological point of view we know that the occurrence of reticulate evolution is neither independent nor distributed uniformly. For example in some species, hybridization plays an important role in evolutionary biology. We know that plants hybridize much more readily than multicelled animals. American oaks are a well-known example (Van Valen, 1976). Also horizontal gene transfer (HGT) is a major factor in bacterial evolution and often occurs between parasites and their hosts (Nickrent et al., 2004). In contrast, the process of horizontal gene transfer is not common to all species. Based on these biological facts one may reasonably raise some doubt about the validity of the second criteria. For purposes of this paper, we consider $R(N)$ to be the most significant criteria for complexity of a network.

Before explaining our method, it will be useful to comment briefly on why non-dense triplet sets are so important.

Why non-dense?

As mentioned in the first part of this introduction, two motivations behind using triplets are the accuracy and the data disparity problem. Obviously, if we restrict ourselves to dense triplet sets as input, then we can not overcome the data disparity problem. Moreover, another reason for using non-dense data sets is that we are able to construct a very accurate input from every character data because we are allowed to choose the most accurate triplets by applying threshold criteria. Clearly, this is not possible for dense triplet sets.

## Methods

In this section we will describe the TripNet algorithm. Before we discuss the different steps of the algorithm in detail, we first give an overview of the definitions we use in the remainder of the paper and then explain the motivation of our approach through an example.

Definitions and Notation

A phylogenetic network is a rooted directed acyclic graph in which every node except the root satisfies one of the following conditions:
a) It has indegree 2 and outdegree 1. These nodes are called reticulation nodes.
b) It has indegree 1 and outdegree 2.
c) It has indegree 1 and outdegree 0. These nodes are called leaves.

A triplet is a binary rooted tree with three leaves. We use $ij|k$ to denote the triplet with taxa $i$ and $j$ on one side of the root and $k$ on the other side of the root (see Figure 1(a)). A triplet $ij|k$ is consistent with a phylogenetic network $N$ if $N$ contains a subdivision of $ij|k$. Figure 1(b) shows an example of a network and a triplet consistent with it. A set $\tau$ of triplets is consistent with a phylogenetic network $N$ if all the triplets in $\tau$ are consistent with $N$.

As stated in the introduction, given a set of triplets, TripTree decides in polynomial time whether there exists a tree consistent with all triplets. In the case that there exists such a tree, the algorithm produces it. Now if there is no tree consistent with a set of triplets $\tau$, we should produce a phylogenetic network $N$ consistent with $\tau$, as close to a tree as possible. As stated in the introduction, generating an optimum phylogenetic network which is compatible with a set of triplets is a hard problem. Intuitively, one of the difficulties of solving this problem is the lack of suf-ficient information for detecting the structure of the optimum network. Thus, using the high density condition, we can solve the problem in polynomial time in some situations (To and Habib, 2009).

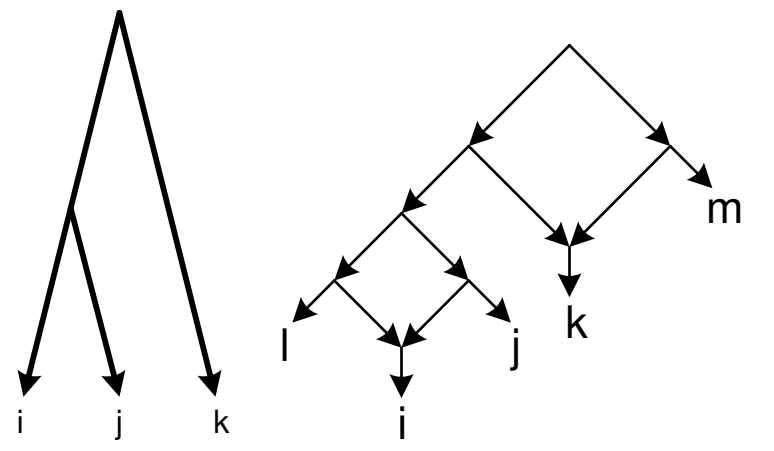

(a) $i\ j|k$ (b) $N$

FIG. 1.—$i\ j|k$ is consistent with $N$

The basic idea of the TripNet algorithm is to find a height function as an intermediate computational step which yields the minimum amount of information required to construct the network. This approach is inspired by ideas from Morse theory, which is well known in singularity theory for its capacity to characterize the topology of surfaces. We use the height function for detecting the complexities of the desired network.

The height function has more information compared to the set of triplets. Before illustrating the TripNet algorithm in detail, we first describe TripTree algorithm based on the height function to show how it can be used to construct a tree from triplets. Then, an example of using the height function for reconstruction of networks is presented.

Assume $T$ is a binary tree and $h$ is a function which assigns to each pair of leaves $x$ and $y$ in $T$, the height of the lowest common ancestor of $x$ and $y$. Define a weighted complete graph $(G,h)$ where $V(G)$ is the set of leaves of $T$ and each edge $ij$ has weight $h(i,j)$.

First, remove the edges with maximum weight from $G$. Obviously this operation will disconnect the leaves of the left subtree from the leaves of the right subtree in $G$. In fact this operation corresponds to removing the root from $T$. The process of removing the edges with maximum weight is continued in the connected components. In each step the leaves of two subtrees will be separated and each step can be seen as removing the highest nodes of remaining subtrees. At the end of this procedure one can easily reconstruct the tree by reversing the steps of the algorithm which is depicted in Figure 2.

Now if a function $h$, which assigns a nonnegative integer to each pair of taxa, is given as input, the algorithm above can decide in polynomial time whether a tree with height function $h$ exists. In fact such tree exists if and only if in all steps, removing the edges with maximum weight causes the graph to become disconnected. This is exactly the TripTree algorithm.

In the TripNet algorithm, if a tree exists, the method of generating the tree is the same as TripTree. It should be noted that, if in one step, by removing the edges with maximum weight the graph is still connected, TripNet is able to continue for detecting the structure of the network. For detecting the structure of the network, the process of removing the edges with maximum weight is continued until the specific component becomes disconnected. The network of Figure 3 presents this opportunity. The existence of such a situation means the existence of at least one reticulation node in the network. Evidently, the reconstruction of the network after the end of this procedure is not as simple as the reconstruction of the tree. In addition, there is not a simple method for creating the height function. The TripNet algorithm proposes solutions for these problems.

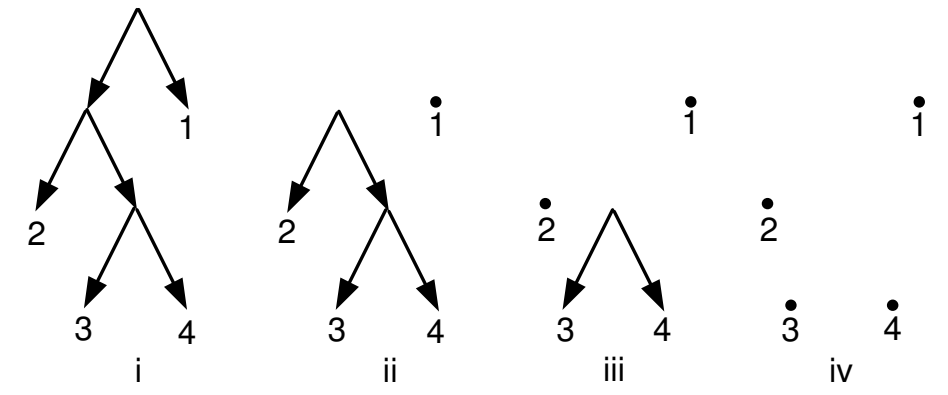

FIG. 2.—The process of removing highest node in each tree and resulting subtrees. $h(1,2) = h(1,3) = h(1,4) = 3$, $h(2,3) = h(2,4) = 2$, $h(3,4) = 1$

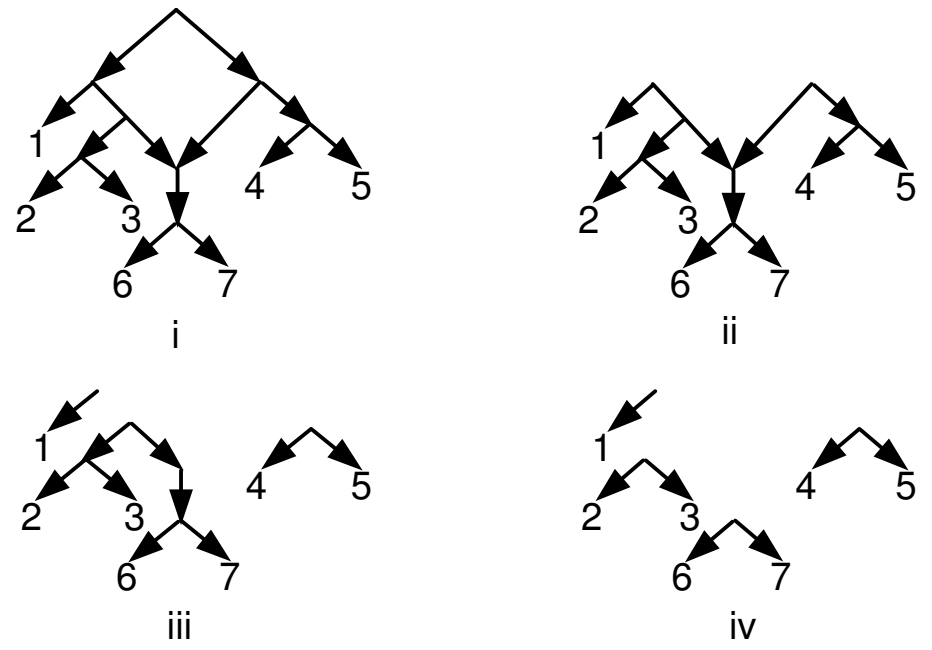

FIG. 3.—The steps of removing edges with maximum weight from the network

Lowest common ancestor of two leaves is not a well-defined concept in networks because there may be more than one lowest common ancestor for two leaves. Intuitively, $h(i,j)$ is the height of one of the lowest common ancestors for i and j in the desired network. We assume that the height of all leaves is zero and the root is the highest node. For example suppose $i$, $j$ is a cherry (a pair of leaves that are adjacent to a common node). Then $h(i,j) = 1$. Obviously every network $N$ indicates a unique height function $h_N$ but for a given height function $h$ there are many networks $N$ such that $h_N = h$. We say that such networks are consistent with $h$.

Let $L$ be the set of all leaves that appear in at least one of the triplets in $\tau$. A subset $S$ of the leaves is an $SN$-set if there is no triplet $ij|k \in \tau$ such that $i \notin S$ and $j,k \in S$.

By finding the $SN$-sets and contracting each of them to a node, we assume a common ancestor for all of these leaves. Note that no triplet in the form of $ij|k$ which $i$ and $j$ are in one $SN$-set and $k$ is not, exists. Thus, the final constructed network is compatible with all of the triplets.

Therefore, contracting the *SN*-sets to one node reduce the size of the problem. This method is discussed in the papers related to the reconstruction of phylogenetic networks based on dense triplets (Jansson and Sung, 2006). In these papers, for finding the *SN*-sets in polynomial time, the authors use the high density of the input triplet sets. In the TripNet algorithm, the high density assumption is removed by using the concept of height function.

Now we describe TripNet in eight steps and illustrate the steps by example in Figure 4.

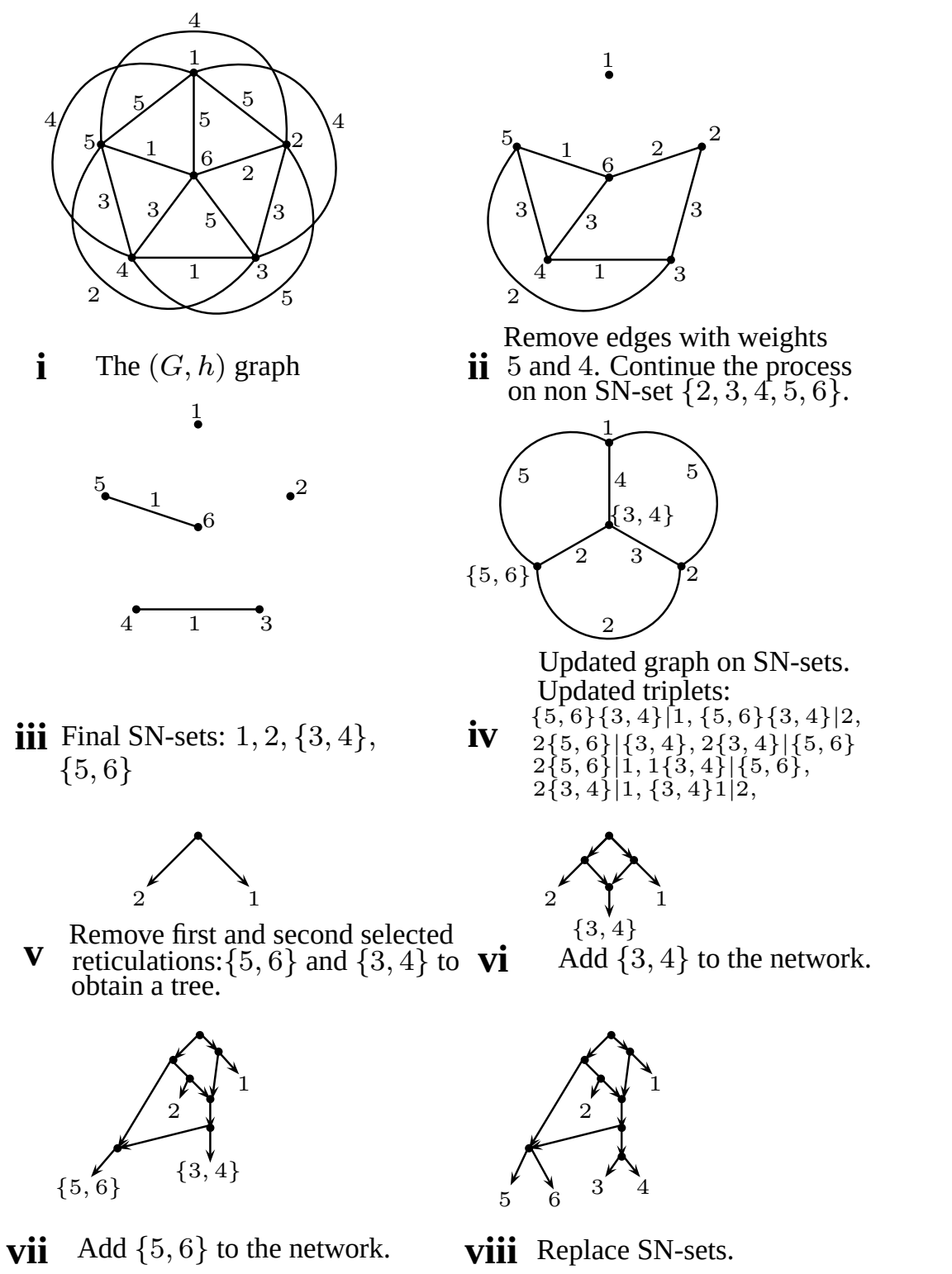

FIG. 4.—Steps of TripNet algorithm for input triplets: 2 3|1, 4 1|2, 5 2|1, 2 6|1, 3 4|1, 1 3|5, 1 3|6, 5 4|1, 4 6|1, 5 6|1, 3 4|2, 3 5|2, 2 6|3, 4 5|2, 2 4|6, 5 6|2, 3 4|5, 3 4|6, 5 6|3, 5 6|4

**Step 1** : In this step we want to find a height function $h$ from a given triplet such that every network which is consistent with $h$ is consistent with the given triplets. The main idea behind the computation of the height function comes from the following simple observations:

1. If in a network $N$, $i$, $j$ have a common ancestor in a lower height compared with a common ancestor of $i$, $k$ (or $j$, $k$) then reasonably we expect that $ij|k$ is a consistent triplet with $N$ and vice versa. So for every triplet $ij|k$ we expect that $h(i,j) < h(i,k)$ and $h(i,j) < h(j,k)$.

2. If the number of taxa is $n$ then the number of pairs $i$, $j$ is $c(n,2)$ and there is a height function $h$ consistent with all given triplets such that for every pair $ij$, $0 < h(i,j) \leqslant c(n,2)$.

3. Intuitively, among networks consistent with a given height function $h$, the networks with highest root have less reticulation nodes.

So we must solve the following IP (Integer Programming):

$$\begin{array}{ll} \text{Maximize} & \Sigma_{i,j} h(i,j) \\ \text{Subject to} & h_{(i,k)} - h_{(i,j)} \geqslant 1 \\ & h_{(j,k)} - h_{(i,j)} \geqslant 1 \\ & 0 < h_{(i,j)} \leqslant c(n,2) \end{array}$$

Fortunately, there is a simple combinatorial method to find the solution of the IP above.

Define a directed graph $G'$ by $V(G') = \{ij|i,j \in L\}$ $E(G') = \{\ (ij,ik),\ (ij,jk) \mid ij|k \in \tau\}$ (The nodes $ij$ and $ji$ are the same ).

The IP above has a feasible solution if and only if the graph $G'$ is a directed acyclic graph $(DAG)$. If $G'$ is not a $DAG$ we remove some edges from $G'$ in such a way that the resulting graph $G''$ is a $DAG$. Since every edge in $G'$ corresponds to some triplet, by removing an edge from $G'$ we are omitting the effect of the corresponding triplet. However, any such missing information will be recaptured in step 8. Now in $G''$ if $l_m$ denotes the node length of the longest path, assign $l_m$ to the nodes with outdegree zero and remove them. assign $l_m - 1$ to the nodes with outdegree zero and continue until all nodes are removed. Then the numbers assigned to nodes is a solution to the IP above in which some inequalities are removed.

**Step 2** : Define a weighted complete graph $(G,h)$ where $V(G)$ is the set of taxa and each edge $ij$ has weight $h(i,j)$.

**Step 3** : Remove the maximum-weight edges from $G$. Continue the process of removing edges until the resulting graph is disconnected.

**Step 4** : Contract each connected component which is an *SN*-set to a single node. If a connected component is not an *SN*-set, continue the process of removing maximum-weight edges until it becomes disconnected and continue the process of removing edges until all of the connected components become *SN*-sets. Now update the set of triplets with respect to *SN*-sets. The updating process is done by the following procedure: Let $c_1$, $c_2$, ..., $c_k$ be the *SN*-sets. Replace $\tau$ by $\tau_c = \{c_i\, c_j | c_k \mid ij|k \in \tau, i \in c_i, j \in c_j, k \in c_k\}$.

**Step 5** : In this step the reticulation nodes are recognized using some heuristic criteria. Let $mc_i$ and $Mc_i$ be the minimum and maximum weight of edges in $(G,h)$ with one end in $c_i$. For example the first criteria is to choose the node with minimum $mc_i$ and if there is more than one node with minimum $mc_i$ then choose the node with minimum $Mc_i$. Other criteria have been explained in detail in the pseudocode. If more than one node passed all criteria then we choose the reticulation node by trying all or some of the remaining candidate nodes based on which mode of TripNet is running (slow, normal, or fast). By deleting the reticulation nodes, the resulting network is a tree. We construct this tree and add the reticulations to it.

**Step 6** : For each *SN*-set and the set of triplets which all of its taxa are in the *SN*-set we run the algorithm again.

**Step 7** : We replace each *SN*-set in network of step 5 with its related network constructed in step 6.

**Step 8** : In this step we check whether the constructed network is consistent with all input triplets. If not we add some reticulation nodes to justify the consistency of all triplets.

## Results

In this section we illustrate the application of the method described in section 2 to the problem of reconstructing phylogenetic networks from triplets. We performed two simulations and also reanalyzed five published data sets using TripNet. The data sets and TripNet are available online from the TripNet webpage.

### Simulations

In the following, we will present two simulation results which show the performance of TripNet under two scenarios. In the first scenario, 50 different phylogenetic networks are randomly generated by generating random binary trees and adding random edges to make reticulation nodes. The total number of reticulation nodes in the generated networks is 200 and each network has at least 1 and at most 10 reticulation nodes. Then, for each network $N$, the set of all triplets consistent with $N$ is computed. Finally, each of these triplet sets (or a random subset of them) is used as the input to TripNet with the hope that the resulting networks are essentially the same with not much additional reticulation nodes. The result of the experiments showed that the resulting networks essentially preserve the structure of initial networks. The average, minimum and maximum of the difference between the number of reticulation nodes in TripNet networks and original networks are shown in Table 1.

| loss percent | 0 | 0.2 | 0.4 | 0.6 | 0.8 |
|---|---|---|---|---|---|
| Min | 0 | 0 | 0 | 0 | 0 |
| Max | 5 | 5 | 7 | 4 | 4 |
| Avg | 1.2 | 1.1 | 1.48 | 1.48 | 1 |

Table 1 : Simulation results on fifty randomly generated triplet sets

It is worthy to note that in this simulation, in some cases all three possible triplets on three taxa may be included in input sets. In order to study the performance of TripNet in displaying biological reticulation events we design a more realistic scenario. Create a random tree based on the following method. A random DNA sequence of 50 codons was generated at the root of the tree. The sequence is then "evolved" along the branches of the tree by simulating random mutations in random sites with at most one mutation in each site. We obtained a tree with eight leaves which all are labeled by randomly generated DNA sequences. TripNet outputs a tree on these 8 sequences as depicted in Figure 5(a). Then 4 new sequences are created by recombination between every pair of sequences at random sites (except for cherries). Each time one of these new sequences is added to the 8 original sequences. TripNet outputs a network on these 9 sequences. The percent of the networks which have 0, 1, 2, and 3 reticulation nodes are 13 %, 69 %, 15 %, and 3 % respectively. For each case one output is depicted in Figure 5. We can conclude that TripNet can reasonably detect and represent the recombination events.

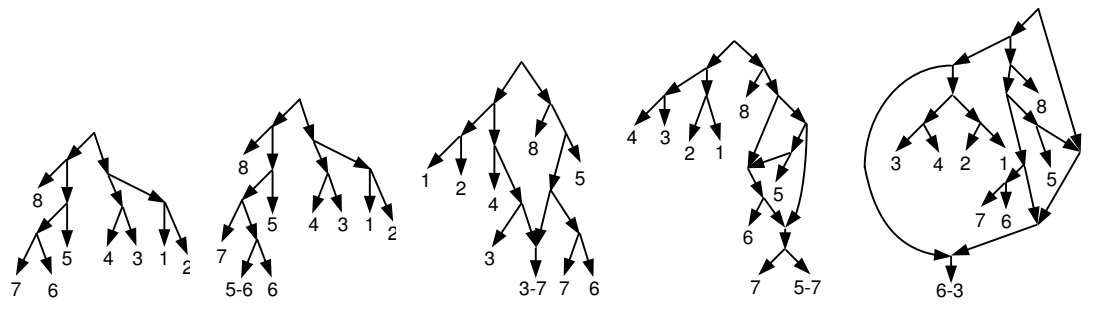

(a) (b) (c) (d) (e)

FIG. 5.— Randomly generated tree and recombination networks

### Kreitman data

In Song and Hein (2003), a method is presented which given an arbitrary set of haplotypes as input, outputs a network with minimum recombination. For "Kreitman's 1983 data of the alcohol dehydrogenase locus from 11 chromosomes of Drosophila melanogaster", a network is presented which contains 7 reticulation nodes. For this data, three haplotype sequences are the same and thus the final network has 9 taxa (Figure 6(a) ).
In our experiment, using the DOLLOP program from the PHYLIP package, for each subset of three haplotype sequences, the most probable trees based on the Parsimony criterion is produced. (In this method, for some of the subsets, two or three trees are presented and all of them are included in the set of triplets.) For this set of triplets, TripNet produced a network with 5 reticulation nodes with a structure similar to the network which is produced by the above algorithm (Figure 6(b) ).

The number of reticulation nodes should be less in the TripNet network because there is no restriction that the kind of the reticulate event be a recombination.

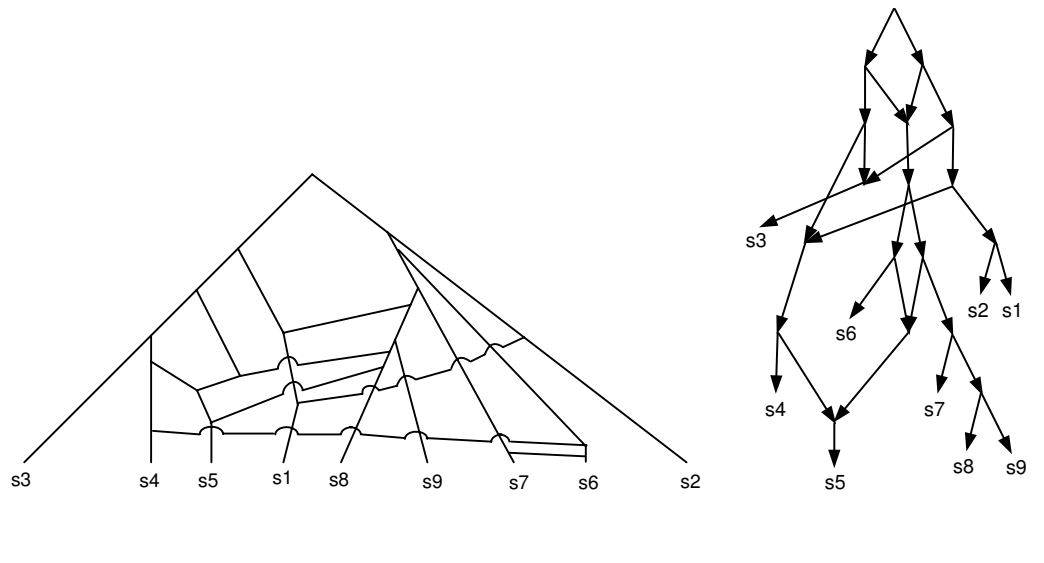

(a) Recombination method (b) TripNet

FIG. 6.—Resulting networks from Kreitman haplotypes

### Yeast Data

The Yeast data is a dense set of rooted triplets generated using real yeast data, obtained from the Fungal Biodiversity Center in Utrecht. This data set which contains information about 21 species is available online from (http://homepages.cwi.nl/∼kelk/level2triplets.html). Based on the algorithm developed in Iersel et al. (2008). Steven Kelk has developed a software application, called LEVEL2, for constructing level-2 phylogenetic networks from dense sets of triplets. LEVEL2 is not applicable to general triplet sets and it produces a network only if there exists a level-2 phylogenetic network consistent with the input triplets. However, LEVEL2 has the advantage that it always produces the best possible network. LEVEL2 network for the Yeast data is a 21-leaf level-2 network which is given in Figure 7(a).

As our only chance for comparing TripNet networks with the best possible networks we repeated the analysis of Yeast data using TripNet. The TripNet network for the Yeast dataset is given in Figure 7(b).

As one can see, TripNet produced a level-3 network which contains only one more reticulation than the network obtained by LEVEL2. Both networks have the same clustering and represent the same evolutionary relationship between taxa. While TripNet has been designed for general triplet sets (not necessarily dense or consistent with a restricted level network), this example shows that the network produced by TripNet is very close to the best possible solution.

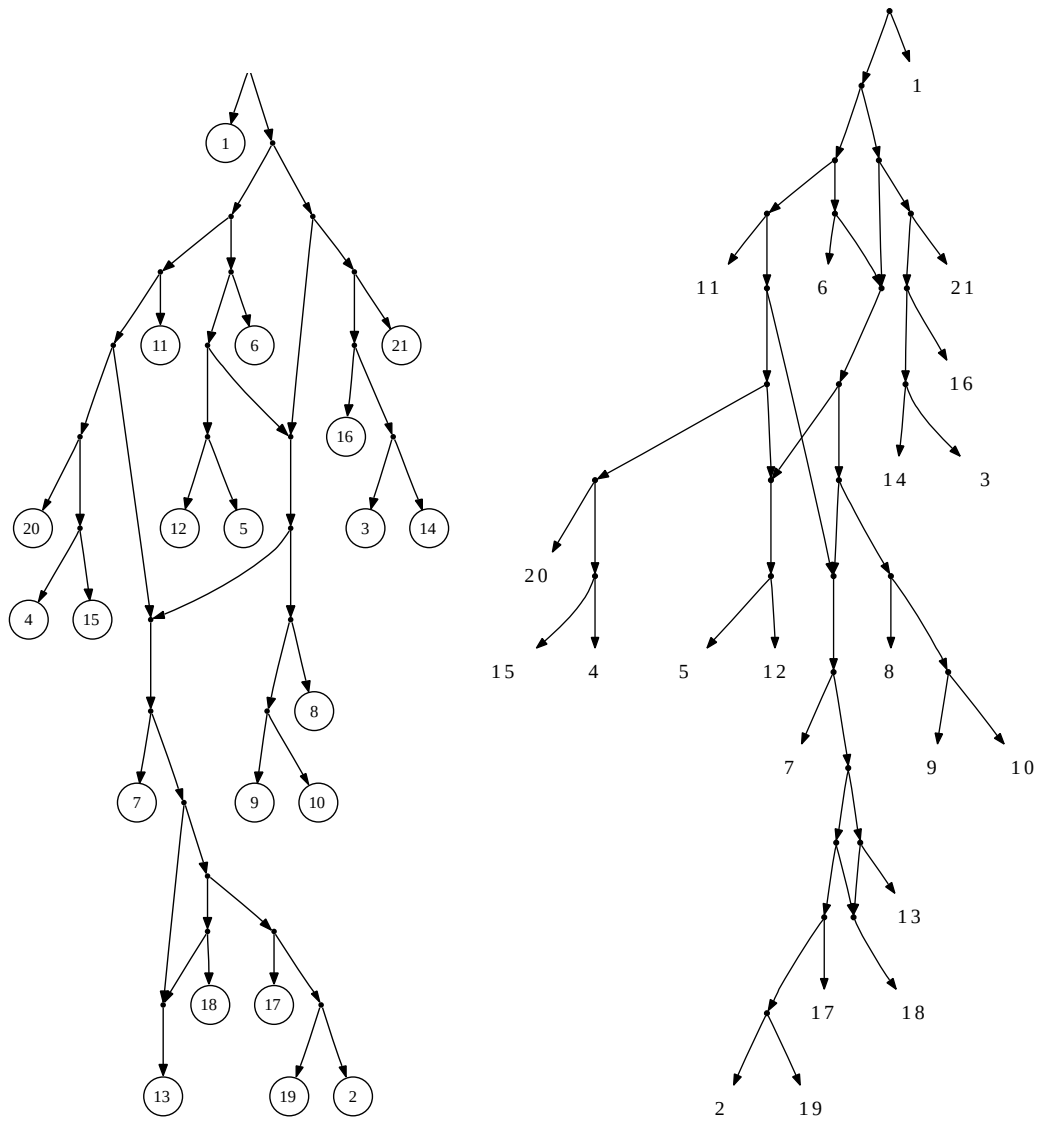

(a) LEVEL2 network (b) TripNet network

FIG. 7.—Resulting networks from Yeast triplets

## Salmonella MLST Data

Similar to Bryant and Moulton (2004) and Grunewald et al. (2007) we re-analyzed the 110 phosphomannomutase (manB) sequences, published in Kotetishvili et al. (2002) and compare the TripNet Network for this data with those from Neighbor-Net and QNet.

The resulting networks by Neighbor-Net and TripNet for the complete set of 110 sequences are depicted in Figure 8(a) and 8(b). Clearly, both algorithms produced the same major clusters. As expected both networks indicate complex patterns of evolution.

In Grunewald et al. (2007) the authors have reported that "although the split with one part consisting of Sha146, Sha135, UND8, Sjo99, Sha151 and Sty19* occurs in 93% of the trees constructed by Tree-Puzzle, it is only represented in the QNet network in Figure 8(c) and that, moreover, this network represents also a conflicting split of higher weight. This indicates that trying to represent data by a tree can result in high support for splits that are not supported if the restriction to trees is omitted." On the other hand, in the network obtained by TripNet, these 6 taxa have formed a split which indicates no reticulate pattern between them.

In Bryant and Moulton (2004) the authors performed the sliding-window analysis to test recombination. Since this technique requires a huge amount of computation for the complete set of 110 sequences, instead they used Neighbor-Net to select a small set of seven taxa to test for recombination in a specific area in the network depicted in Figure 9.

They concluded that much of the conflicting signal comes from sites 110-250. The results of their partitioned analysis for these taxa is shown in Figure 9. We also obtained the TripNet network for these seven taxa from different sites.

From TripNet networks depicted in Figure 10 we see that by removing the sites 110-250 the TripNet network is tree-like and the network obtained from sites 110-250 has one reticulation. This shows that the conflicting signal comes from these sites as sliding window suggests. It is worth noting that while TripNet network represents these conflicting signals by only one reticulation, the Neighbor-Net network uses three parallel edges to represent them.

## ITS and JSA data

ITS and JSA are data from New Zealand alpine buttercups (Ranunculus) which has been recently analyzed using Neighbor-Net.

In Lockhart et al. (2001) the authors published a phylogenetic analysis of nrDNA ITS and cpDNA JSA sequence data of New Zealand alpine buttercups of Ranunculus sect. Pseudadonis, using quartet puzzling and split decomposition. This study has provided important new insights into the relationships and evolution of these species. The phylogenetic trees constructed by quartet puzzling indicate that the alpine Ranunculi of New Zealand comprises of four distinct phylogenetic groups. The data for groups I and II were examined in detail under split decomposition. The splits graph of JSA sequences for taxa of group I is tree like but the other three splits graphs are reticulate networks. Now by TripNet we produced two networks for ITS and JSA which each contains 48 taxa (Figures 13 and 14 )

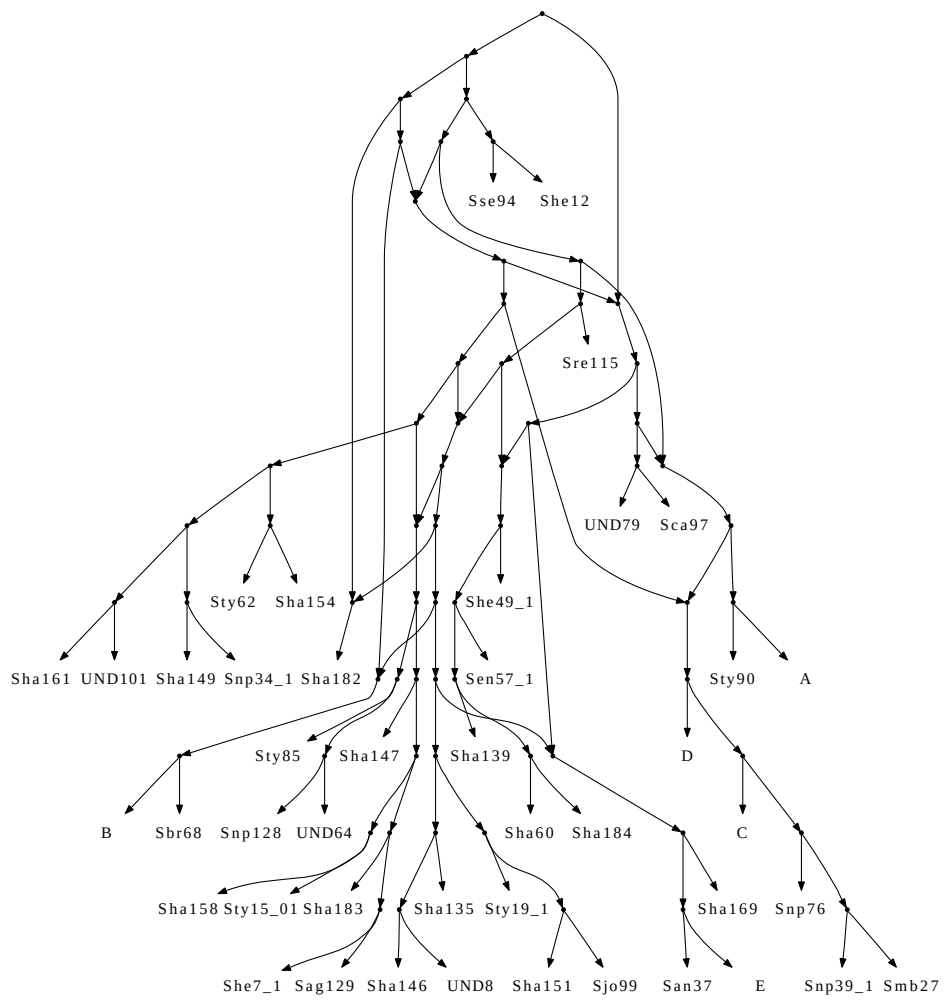

(a) TripNet

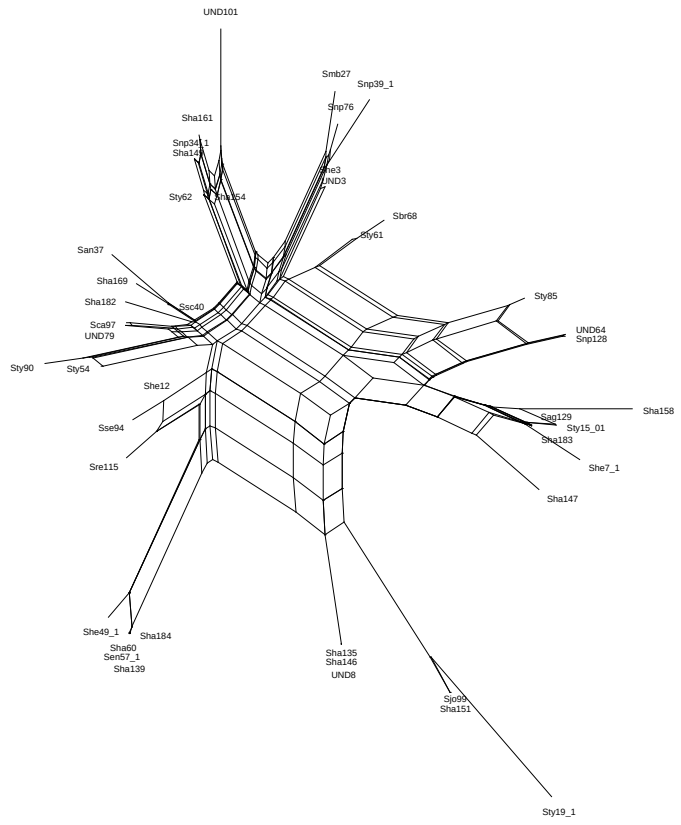

(b) Neighbor-Net

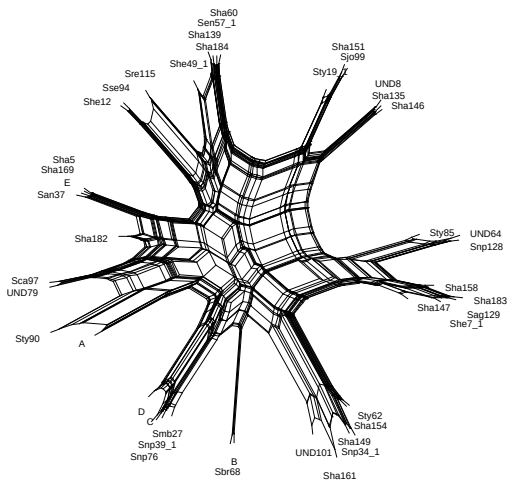

(c) QNet

FIG. 8.—Salmonella

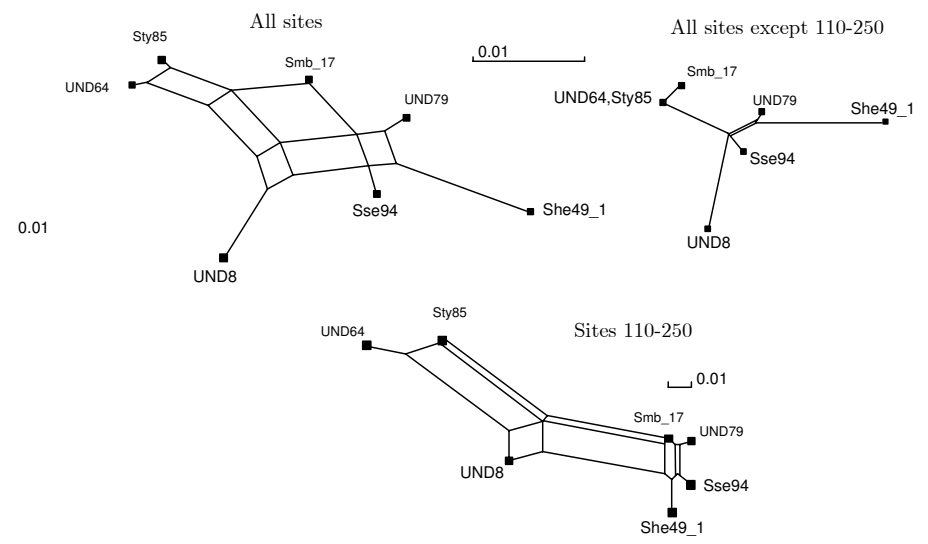

FIG. 9.—Neighbor-Net results on seven taxa

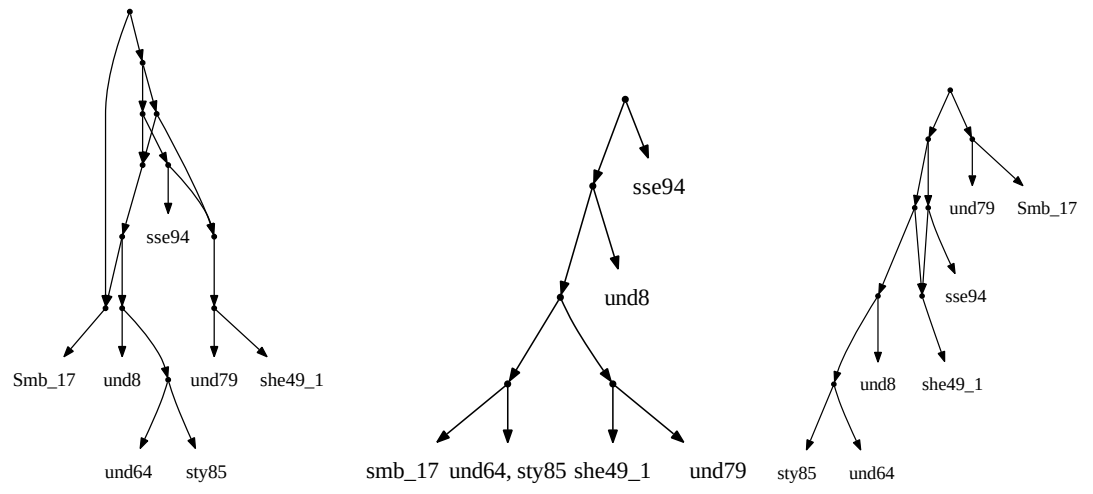

(a) All sites

(b) All sites except 110-250

(c) Sites 110-250

FIG. 10.—TripNet results on seven taxa

## Discussion

In this paper we have developed and implemented a new algorithm, called TripNet, which constructs an explicit phylogenetic network consistent with a given set of triplets. TripNet is an unsupervised algorithm and like SplitsTree, automatically draws the output network. The source code of TripNet is in Java language. Unlike previous methods which only work on dense triplet sets our method works on any set of triplets, so we are able to overcome the data disparity problem and constructing a very accurate input from every character data.
An advantage of triplet-based methods is that the triplets in the input are rooted, and hence, the output network will be

. For both data, the TripNet networks have the same clusters as obtained using tree puzzling. In fact the groups in our networks are comparable with the networks in Fisher (1965) and Lockhart et al. (2001). The TripNet networks for groups I and II in both ITS and JSA are depicted in Figures 11 and 12. These networks are very similar to the networks obtained by Lockhart et al. using split decomposition methods. TripNet produced two networks for both ITS groups I and II. For JSA group I the TripNet produces a tree similar to split decomposition method, but for JSA group II TripNet produces a tree while split decomposition returns a network. But in all four cases the relation between taxa are similar.

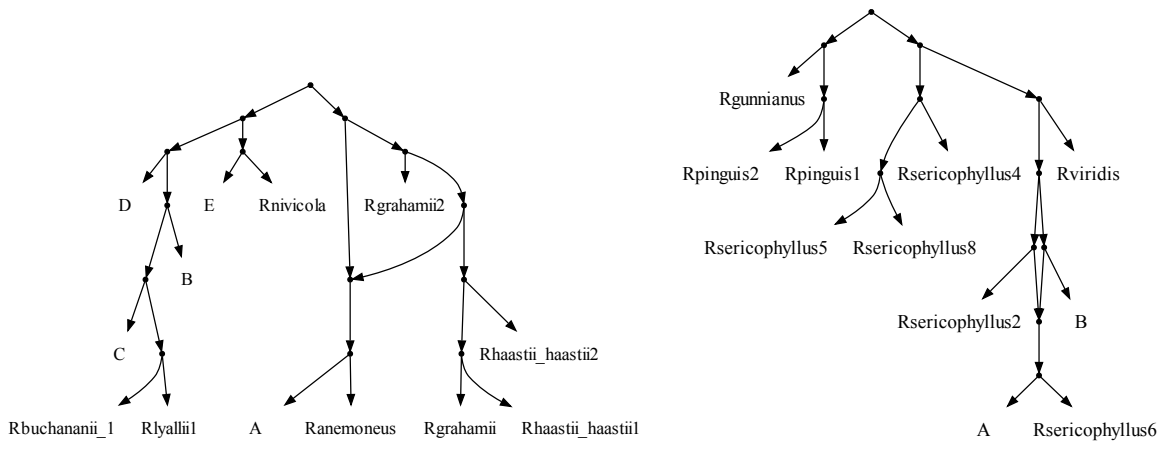

(a) its48group1

(b) its48group2

FIG. 11.—its48group1: *A*={*R. lyallii2, R. lyalliFJ, R. lyalliTB, R. lyallii3*}, *B*={*R. lyallii4, R. lyallii5*}, *C*={*R. buchananii-4, R. buchananii-2*}, *D*={*R. haastii-piliferus1, R. haastii-piliferus2*}, *E*={*R. verticillatus1, R. verticillatus2*} *its48group2: A*={*R. sericophyllus1, R. sericophyllus3*}, *B*={*R. pachyrhizus1, R. pachyrhizus2*}

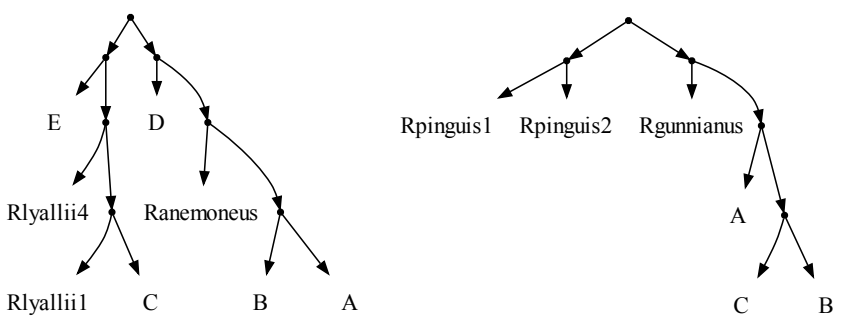

(a) jsa48group1 (b) jsa48group2

FIG. 12.—jsa48group1: *A={R. lyallii2, R. lyalliTB}, B={R. lyalliFJ, R. lyallii3}, C={R. buchananii-4, R. buchananii-1, R. buchananii-2, R. lyallii5}, D={R. haastii-haastii1, R. haastii-haastii2, R. grahamii, R. grahamii2}, E={R. haastii-piliferus1, R. haastii-piliferus2} jsa48group2: A={ R. sericophyllus1, R. sericophyllus2, R. sericophyllus3, R. sericophyllus6, R. viridis}, B={R. sericophyllus4, R. sericophyllus5, R. sericophyllus8}, C={R. pachyrhizus1, R. pachyrhizus2}*

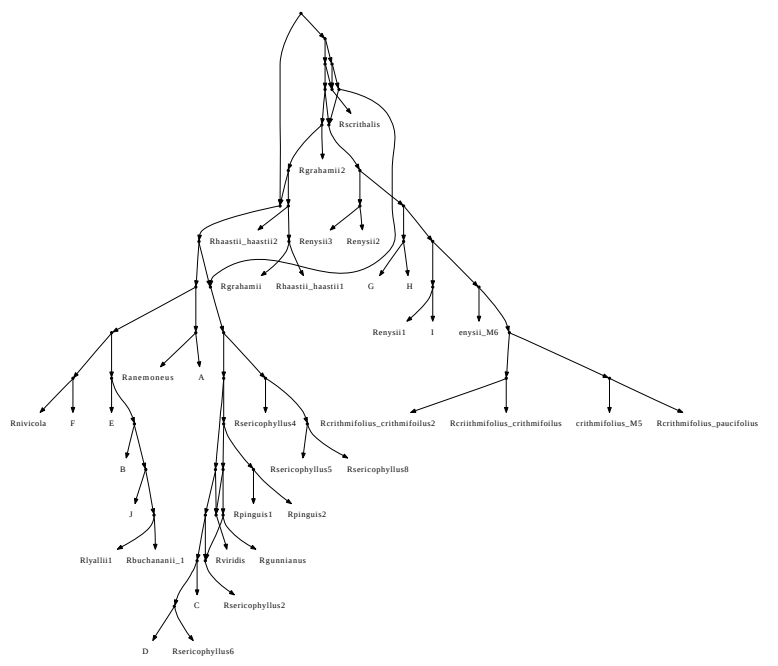

FIG. 13.—ITS48 *A={R. lyallii2, R. lyalliFJ, R. lyalliTB, R. lyallii3}, B={R. lyallii4, R. lyallii5}, C={R. sericophyllus1, R. sericophyllus3}, D={R. pachyrhizus1, R. pachyrhizus2}, E={R. haastii-piliferus1, R. haastii-piliferus2}, F={R. verticillatus1, R. verticillatus2}, G={R. insignis1, R. insignis2}, H={R. godleyanus1, R. godleyanus2}, I={R. gracilipes1, R. gracilipes2}, J={R. buchananii-4, R. buchananii-2}*

rooted, automatically. The distance-based methods such as NNet use a distance matrix as the input and produce an unrooted tree. Similarly, quartet-based methods such as QNet use ( unrooted) quartets as the input and produce an unrooted network. For both of these methods, there often is a final step to make the output rooted. This step may introduce additional errors to the algorithm, or lose some information in the output. TripNet, which produces a rooted tree by nature, does not need this last step.

TripNet is an extension of TripTree, an algorithm developed in Aho et al. (1981). So if there exists a tree consistent with the set of triplets then TripNet will produce a tree. Unlike Aho's algorithm, TripNet can represent conflicting signals in the data. These conflicting signals come from biological sources or sampling errors. The TripNet network generally gives a clear indication of which parts of the network the complexity stems from, allowing us to focus in on those regions, with more detailed and computationally demanding methods. TripNet is able to construct networks with many taxa.

To establish the performance of TripNet we tested it on these five real datasets and two simulation data. Real datasets are Yeast, Salmonella, Its48, Jsa48, and Kreitman data. We compared our results with those of LEVEL-2, Neighbor-Net, QNet, and Song-Hein method Song and Hein (2003) on these five real datasets. The comparisons show that our algorithm is informative and highly accurate and truly determines the evolutionary relationship between species.

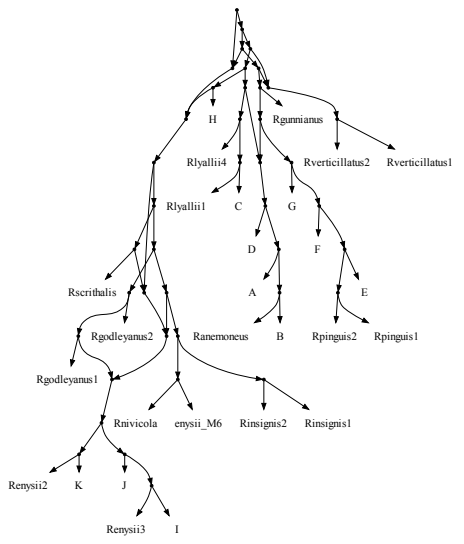

FIG. 14.— JSa48 *A={R. lyallii2, R. lyalliiTB}, B={R. lyalliFJ, R. lyallii3}, C={R. buchananii-4, R. buchananii-1, R. buchananii-2, R. lyallii5}, D={R. haastii-haastii1, R. haastii-haastii2, R. grahamii, R. grahamii2}, E={R. sericophyllus1, R. sericophyllus2, R. sericophyllus3, R. sericophyllus6, R. viridis}, F={R. sericophyllus4, R. sericophyllus5, R. sericophyllus8}, G={R. pachyrhizus1, R. pachyrhizus2}, H={R. haastii-piliferus1, R. haastii-piliferus2}, I={R. criithmifolius-crithmifoilus, R. crithmifolius-crithmifoilus2}, J={R. crithmifolius-paucifolius, crithmifolius-M5}, K={R. gracilipes1, R. gracilipes2, R. enysii1}*

In order to show the running time of TripNet we ran it on these five real datasets on a PC with an Intel Core i7 processor running at 2.80 GHz.

The running time for Kreitman, Yeast, Salmonella, Its48, and Jsa48 data for the almost dense triplets are 1, 1.1, 30, 3.5, and 3.2 second respectively. For Its48, Jsa48, and salmonella usually removing triplets, based on their weight (these weights are obtained from the standard triplets construction methods), cause a decrease in the number of reticulations and the running time for the final network. Another test for TripNet is using simulated data sets. First, we generated fifty networks and all of its consistent triplets randomly. Then we remove some of their triplets randomly. On average TripNet constructs networks with the number of reticulation nodes near to the optimum network. Secondly we showed that if there is a tree for some given species and a new species is obtained from the recombination between two particular species, in the most cases TripNet can show recombination by adding reticulation node(s) to the original tree without disrupting the structure of the tree. In this paper we used standard methods to convert a set of given sequences data into a set of triplets. For the future works, analyzing the methods of converting sequences into triplets are of interest and finding effective methods for generating triplets from sequences data is the most important part of the future works. If one can find a new method for generating reliable set of triplets for some species, TripNet can find new reticulate evolution events between species. Also some parts of the TripNet algorithm are heuristic and can be improved in order to obtain better networks.

**Acknowledgments**

We thank M.sadeghi, A.Asadpour, O.Etesami, and M.Kargar for their kind and useful comments. This work was partially supported by a grant from IPM.

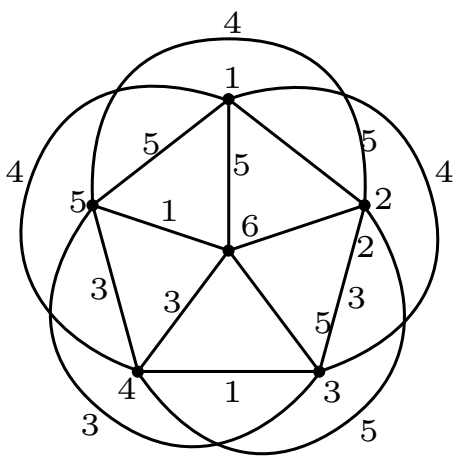

**i** The $(G, h)$ graph

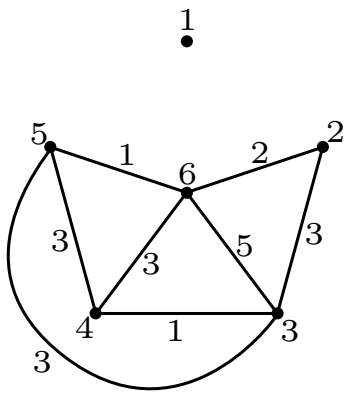

**ii** remove edges with weights $4, 5$; Continue the process on non SN-set component $\{2, 3, 4, 5, 6\}$.

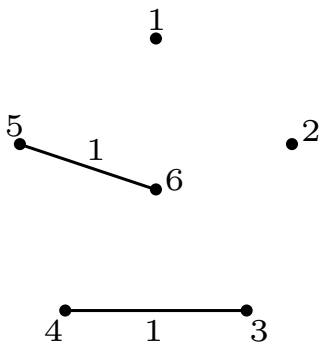

**iii** SN-sets: $1, 2, \{3, 4\}, \{5, 6\}$

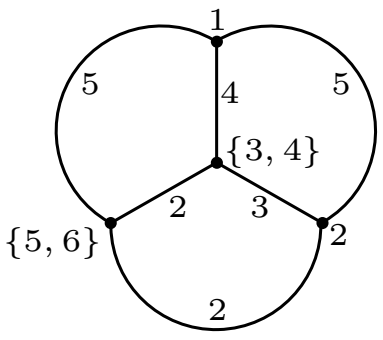

**iv** Contracted component's graph
Updated triplets:
$2\{3,4\}|1$, $\{3,4\}1|2$,
$2\{5,6\}|1$, $1\{3,4\}|\{5,6\}$,
$\{5,6\}\{3,4\}|1$, $\{5,6\}\{3,4\}|2$,
$2\{5,6\}|\{3,4\}$, $2\{3,4\}|\{5,6\}$

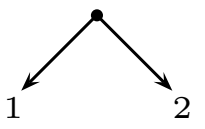

**v** remove first and second selected reticulations: $(\{5, 6\}, \{3, 4\})$, to obtain a tree.

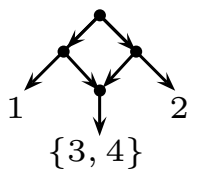

**vi** Add $\{3, 4\}$ to network.

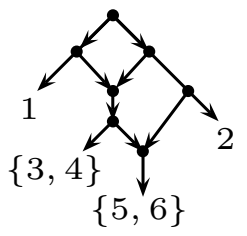

**vii** Add $\{5, 6\}$ to network, based on its calculated right and left neighbors.

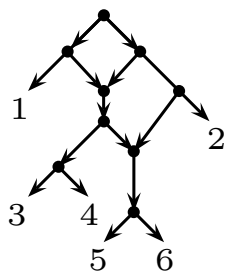

**viii** Replace SN-sets.